\documentclass[3p,amsmath,amssymb,compress]{elsarticle}
\usepackage{amsmath}
\usepackage{graphicx}
\usepackage{epsfig}
\usepackage{latexsym}
\usepackage{amsmath,amssymb}
\usepackage{bm} 
\usepackage{xcolor}
\usepackage{stackrel}
\usepackage{subcaption}
\usepackage{accents}
\usepackage{ulem}
\usepackage{bbm}
\usepackage{comment}
\newcommand{\bp}{\mathbf p}

\newcommand{\br}{\mathbf r}

\newcommand{\dg}{^\dagger}

\journal{Annals of Physics special issue dedicated to the memory of Igor Dzyaloshinskii}

\numberwithin{equation}{section}

\begin{document}
\begin{frontmatter}	

\title{Impurity bands in magnetic superconductors with spin density wave}

\author[Kent]{Maxim Dzero}
\author[UW]{Alex Levchenko}

\address[Kent]{Department of Physics, Kent State University, Kent, Ohio 44242, USA}
\address[UW]{Department of Physics, University of Wisconsin-Madison, Madison, Wisconsin 53706, USA}

\date{May 31, 2022}

\begin{abstract}
Magnetic superconductors define a broad class of strongly correlated materials in which superconductivity may coexist with either localized or itinerant long-range magnetic order. In this work we consider a multiband model of a disordered magnetic superconductor which realizes coexistence of unconventional superconductivity and a spin-density-wave. We derive an exact $T$-matrix and compute a single particle density of states in this system. In a purely superconducting state the interband scattering potential leads to an appearance of the localized Yu-Shiba-Rusinov bound states. Our main finding is that in the fairly broad swath of the coexistence region superconductivity remains fully gapped despite the presence of the impurity bands. We also discuss the effects of spatial inhomogeneities on the density of states in strongly contaminated superconductors. 
\end{abstract}


\end{frontmatter}

\section{Introduction}  

The magnetic disorder in conventional $s$-wave superconductors induces Cooper pair-breaking effects through the scattering with spin flips and leads to the gapless state in accordance with the Abrikosov-Gor'kov (AG) mechanism \cite{AG}. In their theory AG assumed weak short-range impurity potential that can be described within the self-consistent Born approximation (SCBA). This model results in the prediction of the gradual suppression of the superconducting transition temperature and global gap closing in the density of states at the critical concentration of magnetic impurities. In contrast, strong impurity centers were shown to result in the appearance of subgap energy states that are localized in the vicinity of the individual impurity atoms. To capture this effect one must go beyond the Born approximation and employ full $T$-matrix analysis as demonstrated in the original works of Yu \cite{Yu}, Shiba \cite{Shiba}, and Rusniov \cite{Rusinov}. In this picture the corresponding gap suppression occurs only locally. The connection between these two extreme limiting cases can be understood by considering finite impurity concentration. When the magnetic impurities are brought close to one another, the individual localized Yu-Shiba-Rusinov (YSR) states hybridize
and can form energy bands with hard gaps in the averaged density of states.
In turn, the formation of an intragap bound state and impurity bands due to magnetic impurities leads to filling of the superconducting gap and
therefore connects to the AG theory. Further complications to these scenarios may arise in the more realistic models of disorder potential. For example, Larkin and Ovchinnikov \cite{LO} had shown that system inhomogeneities with large correlation radius, longer than superconducting coherence length, may result in the Lifshitz subgap tail states \cite{Lifshitz} in the spectrum of a superconductor even for nonmagnetic disorder. In particular, this result manifestly violates Anderson theorem \cite{Anderson} that suggests spectral rigidity towards inclusion of the quenched nonmagnetic disorder.     

In addition to the spectral modification and gap suppression, localized magnetic moments may change the nature of superconductivity itself. The most striking recent example can be found in the system of a thin-film superconductor decorated by a linear chain of magnetic adatoms deposited on its surface \cite{Nadj-Perge2013,Klinovaja2013,Pientka2013,Bernevig2016}. In this case the superconducting electrons mediate Ruderman-Kittel-Kasuya-Yoshida (RKKY) interactions between the localized magnetic moments of the chain. By virtue of this interaction, the impurity spins may form periodic helical texture, which in general is incommensurate with the underlying impurity chain, and as the consequence the YSR bands may effectively realize a topological superconducting phase, akin to one-dimensional spinless $p$-wave superconductor \cite{Kitaev2001}. The signature feature of this state is that it supports Majorana bound states localized at the ends of the chain which can be tested by scanning tunneling and Josephson spectroscopy probes \cite{Yazdani2014,Yazdani2016}. This example illustrates highly nontrivial effect of the spin ordering in the presence of superconductivity on its spectral properties.  

Recent experimental advances in a synthesis of novel materials, which exhibit strong correlations between the constituent particles have shown, in particular, that at low temperatures superconductivity often finds itself in competition with either itinerant or local-moment magnetism. For the present study the earlier theoretical works on superconductivity in iron-based superconductors are of special interest (for a recent reviews see e.g. \cite{Chubukov2012,Matsuda2014} and references therein). For these systems it has been demonstrated that in the multiband models with the interband and intraband disorder scattering, there will be a region of coexistence between extended unconventional $s^{\pm}$-wave superconductivity (SC) and spin-density-wave (SDW) orders \cite{VC2011,FVC2012}. The same effect can be achieved in a clean system by varying the anisotropy of the electron- and hole-like Fermi pockets \cite{VVC2010,Schmalian2010}. Thus, these models provide a valuable framework for studying the physical consequences of the coexistence between these two canonical and mutually antagonistic long-range orders. The question of whether the phenomenon of gapless superconductivity persists in the coexistence region is one that motivated this study. We note that the observation of gapless superconductivity would in principle allow one unambiguously trace the origin of the coexistence as being driven by disorder as opposed to be driven by the Fermi pockets anisotropy.

To place our work in the context of existing studies we remind that the effect of impurities on the pairing state of unconventional superconductors in general \cite{Balatsky}, and in pnictides in particular, has been widely investigated theoretically (see, e.g., Refs. \cite{Kontani2009,Efremov2011,Vakaryuk2012,Mishra2013,Yamakawa2013,Stanev2014,Hoyer2015,Babaev2017,Brydon2021,Yerin2022}). Experimentally it was revealed that doping as a source of disorder leads to a nonmonotonic evolution of superconducting gaps and electronic densities of states \cite{Hardy2010}, and may result in superconductivity with broken time-reversal symmetry occupying finite domain of the phase diagram inside the dome of global SC state \cite{Grinenko2020}. Furthermore, increasing the impurity scattering may induce topological change of the superconducting gap structure \cite{Mizukami2014,Cho2016}. An impact of disorder is particularly nontrivial in the coexistence region as it may boost superconducting critical temperature \cite{FVC2012}. This is so as for $s^{\pm}$ state it is only the interband scattering that acts as a pair-breaking source, whereas both intra and interband scattering influence SDW order parameter. As a result, SDW suppression has a stronger effect on SC enhancement than the tendency of disorder to suppress it. On top of that pairing-potential disorder leads to a broadening of the coexistence region \cite{Dzero2021}. Thus far these features in disordered pnictide systems were investigated in the corresponding band models within the self-consistent Born approximation. In light of recent experimental findings \cite{Hashimoto,Carrington,Auslaender,Joshi}, this approach was successful in capturing several interesting results concerning the phase diagram and thermodynamic properties \cite{Levchenko2013,Chowdhury2013,Kuzmanovski2014,Dzero2015,Kirmani2019,Fernandes2020,Khodas2020,Hasan2022}. However, it is clear, that this approximation, being perturbative by construction, is not sufficient when one is concerned with the effects -- such as formation of the bound states -- whose description requires one to go beyond perturbation theory. Therefore it is of great interest from both experimental and theoretical points of view to investigate the functional form of the single-particle density of states in the presence of strong disorder in the coexistence region.

The rest of the paper is organized as follows. In Sec. \ref{sec:model} we introduce the two-band disorder model that captures SC-SDW coexistence. We formulate mean-field equations that describe the phase diagram and introduce single-particle propagator that contains properties of the energy spectrum. In Sec. \ref{sec:t-matrix} we introduce an exact $T$-matrix and solve corresponding integral equation in the model of strong short-range impurities. We benchmark the obtained solution against analogous known cases for single and multiple impurities. In Sec. \ref{sec:dos} we promote this treatment to include magnetic SDW order and focus on studying the density of states. We show that superconductivity in the coexistence phase remains fully gapped, i.e. the threshold energy to break the Cooper pairs is always finite throughout the region. Finally, in Sec. \ref{sec:discussion} we briefly discuss expected modifications of obtained results going beyond the mean-field analysis due to optimal disorder fluctuations. In particular, we highlight how random spatially varying parameters of the model, namely coupling constants in the SC and SDW pairing channels, are expected to round off sharp gap features of the impurity band.    


\section{Disorder model of SC-SDW coexistence}\label{sec:model}

In this work we adapt the two-band disorder model introduced earlier in Refs. \cite{VC2011,FVC2012} for the interplay between itinerant SDW and $s^{\pm}$ SC. 
In brief, in this model one considers a circular hole pocket at the center of the Fe-only Brillouin zone, and an elliptical electron pocket displaced from the center by $Q = (\pi,0)$ (or $(0,\pi)$). This model accounts for interactions between the low-energy fermions in the SDW (particle-hole) and SC (particle-particle) channels, as well as their interaction with nonmagnetic impurities. In each interaction channel, the four-fermion term is decoupled via the Hubbard-Stratonovich transformation by introducing SC-$\Delta$ and SDW-$M$ order parameters. This leads to a mean-field theory description.   

The starting point of our analysis is an expression for the single-particle fermionic propagator whose analytical properties contain the information about the energy spectrum 
\begin{equation}\label{Eq0}
\hat{G}(\mathbf{r}_1,\tau_1;\mathbf{r}_2,\tau_2)
=-\left\langle\hat{T}_\tau\{\hat{\Psi}(\br_1,\tau_1)\hat{\Psi}\dg(\br_2,\tau_2)\}\right\rangle.
\end{equation}
Here we introduced the eight-component spinor $\hat{\Psi}(\br,\tau)$ in the Balian-Werthammer representation \cite{BW-PRL63}, which contains spin-$\frac{1}{2}$ $(c,f)$-fermionic fields at point $\br$ and describe two (one electron-like and one hole-like) bands respectively. In a clean limit and in the Fourier momentum and frequency representation the single-particle propagator is of the following form
\begin{equation}\label{Eq1}
\hat{G}_0(\bp,\omega_n)=\frac{-i{\omega}_n{\hat{\tau}_0\hat{\rho}_0\hat{\sigma}_0}+\xi_\bp{\hat{\tau}_3\hat{\rho}_3\hat{\sigma}_0}}{{\omega}_n^2+\xi_\bp^2+\Delta^2+M^2}+\frac{\Delta{\hat{\tau}_3\hat{\rho}_1\hat{\sigma}_0}-M{\hat{\tau}_1\hat{\rho}_0\hat{\sigma}_3}}{{\omega}_n^2+\xi_\bp^2+\Delta^2+M^2}.
\end{equation}
Here $\omega_n=\pi T(2n+1)$ are Matsubara frequencies, $T$ is the temperature, $\xi_\bp=p^2/2m-\mu$ is a single particle dispersion, $\mu$ is a chemical potential. We note that unlike in the corresponding band-model \cite{VVC2010,Schmalian2010}, ellipticity of the Fermi surfaces is not essential for the SC-SDW coexistence in the disorder-model, which is implicit in the simplified form of $\xi_\bp$ in Eq. \eqref{Eq1}.  The pairing amplitudes $\Delta$ and $M$ are accompanied by products of the Pauli matrices $\hat{\tau}_a\hat{\rho}_b\hat{\sigma}_c$ with the subscript $0$ referring to the unit matrix. Each Pauli matrix in this product acts in the band, isospin (i.e. Nambu) and spin subspaces, correspondingly. The order parameters entering into Eq. \eqref{Eq1} must be determined self-consistently via coupled nonlinear integral equations 
\begin{align}\label{Eq2}
&\frac{M}{\lambda_{\textrm{m}}}=-\frac{T}{8}\sum\limits_{\omega_n>0}^{\omega_\Lambda}\int\frac{d^2\bp}{(2\pi)^2}\textrm{Tr}\left[(\hat{\tau}_1+i\hat{\tau}_2)(\hat{\rho}_0+\hat{\rho}_3)\hat{\sigma}_3\hat{G}(\bp,\omega_n)\right], \\ \label{Eq3}
&\frac{\Delta}{\lambda_{\textrm{sc}}}=\frac{T}{8}\sum\limits_{\omega_n>0}^{\omega_\Lambda}\int\frac{d^2\bp}{(2\pi)^2}\textrm{Tr}\left[(\hat{\tau}_0+\hat{\tau}_3)(\hat{\rho}_1+i\hat{\rho}_2)(\hat{\sigma}_0+\hat{\sigma}_3)\hat{G}(\bp,\omega_n)\right].
\end{align}
Here $\omega_\Lambda$ is an ultraviolet cutoff, $\lambda_{\textrm{m}}$ and $\lambda_{\textrm{sc}}$ are the bare interaction constants in the magnetic and superconducting channels, and matrix trace was denoted as $\mathrm{Tr}[\ldots]$. In the clean limit when $\hat{G}=\hat{G}_0$ these equations have only two trivial solutions: $(M=0$, $\Delta=\Delta_0)$ for $\lambda_{\textrm{sc}}>\lambda_{\textrm{m}}$ and $(M=M_0$, $\Delta=0)$ for $\lambda_{\textrm{sc}}<\lambda_{\textrm{m}}$.

As the next step, we introduce time-reversal invariant disorder potential:
\begin{equation}\label{disorder}
\hat{U}(\br)=\sum\limits_{i}\left[u_0{\hat{\tau}_0\hat{\rho}_3\hat{\sigma}_0}+u_\pi {\hat{\tau}_1\hat{\rho}_3\hat{\sigma}_0}\right]\delta(\br-{\mathbf R}_i).
\end{equation} 
The first term here accounts for the intraband scattering, while the second term produces the interband transitions. The sum goes over random locations of impurities labeled by ${\mathbf R}_i$. Then, if we were to ignore the correlations between the impurities, the single-particle propagator averaged over the distribution of disorder with concentration of impurities $n_{\textrm{imp}}$ is
\begin{equation}\label{FullProp}
\hat{G}=\hat{G}_0+n_{\textrm{imp}}\hat{G}_0\hat{\cal T}\hat{G}.
\end{equation}
In this equation the scattering $\hat{\cal T}$-matrix needs to be computed self-consistently. 


\section{Scattering matrix}\label{sec:t-matrix}

The equation for the scattering matrix contains the full propagator
\begin{equation}\label{FullTsEq}
\hat{\cal T}(i\omega_n)=\hat{U}+\pi\nu_F\hat{U}\hat{\cal G}_{\omega_n}\hat{\cal T}(i\omega_n),
\end{equation}
where we introduced the quasiclassical Eilenberger function \cite{Eilenberger}, defined by $\hat{\cal G}_{\omega_n}=\int\hat{G}(\bp,\omega_n){d\xi_\bp}/{\pi}$. In Eq. \eqref{FullTsEq} $\nu_F$ is the normal state quasiparticle density of states at the Fermi energy.

\subsection{Single impurity}

In the case of a single impurity, we can easily compute the scattering matrix by solving Eq. \eqref{FullTsEq} with the propagator taken from Eq. (\ref{Eq1}). We find that the scattering matrix has two pairs of poles (bound states) at energies $\varepsilon_{\textrm{b}}^{(nm)}=(-1)^nc_1+(-1)^mc_2$, ($n,m=1,2$) with parameters $c_{1,2}$ given by
\begin{equation}\label{BoundStates}
c_1=\frac{1+\gamma_0^2-\gamma_\pi^2}{(1+\gamma_0^2-\gamma_\pi^2)^2+4\gamma_\pi^2}
\sqrt{(M^2+\Delta^2)(1-\gamma_0^2+\gamma_\pi^2)^2+4\gamma_0^2\Delta^2}, \quad
c_2=\frac{4\gamma_0\gamma_\pi M}{(1+\gamma_0^2-\gamma_\pi^2)^2+4\gamma_\pi^2},
\end{equation}
where $\gamma_{0,\pi}=\pi\nu_Fu_{0,\pi}$. The parameter $c_2\propto u_0u_\pi$ in the expression for the bound state energy accounts for the interference effects between the intraband and interband scattering processes in a state with nonzero magnetization. A systematic account for such interband coherence goes beyond the scope of this analysis. In fact, the appearance of this term is actually an artifact of the model and, therefore, it will be neglected in what follows.
We further note that in a purely superconducting state, the bound state energy is given by the well-known expression \cite{Shiba,Rusinov}
\begin{equation}\label{ShibaRusinov}
\varepsilon_{\textrm{b}}\vert_{M=0}\approx\pm\left(\frac{1-J_s^2}{1+J_s^2}\right)\Delta, \quad J_s=\frac{\gamma_\pi}{(1+\gamma_0^2)^{1/2}}.
\end{equation}
Indeed, we see that the pair breaking rate is determined by the interband scattering. In the purely SDW state, the bound state energy is
\begin{equation}\label{SDWRusinov}
\varepsilon_{\textrm{b}}\vert_{\Delta=0}\approx\pm\left(\frac{1-\gamma_0^2}{1+\gamma_0^2}\right)M.
\end{equation}
Note that the energy of the bound states in the purely SDW state is much lower than the one in the purely superconducting state, 
$|\varepsilon_{\textrm{b}}\vert_{\Delta=0}\ll|\varepsilon_{\textrm{b}}\vert_{M=0}$.

By solving Eq. (\ref{FullTsEq}) we can in principle compute the full $\hat{\cal T}$-matrix. The general expression, however, is too cumbersome to present here. Instead, we employ the approximation adopted by Rusinov \cite{Rusinov}, which consists of keeping in $\hat{\cal T}$-matrix only those terms that have the same matrix structure as the single particle propagator in Eq. (\ref{Eq1}). This assumption is certainly valid when $\gamma_{0,\pi}\ll1$. We thus find
\begin{equation}\label{CompactDia}
\pi\nu_F\hat{\cal T}(i\omega_n)\approx\sqrt{\omega_n^2+\Delta^2+M^2}\left(\omega_n^2+c_1^2\right)\left[-i\omega_n(\gamma_0^2+\gamma_\pi^2)
{\hat{\tau}_0\hat{\rho}_0\hat{\sigma}_0}-\Delta(\gamma_0^2-\gamma_\pi^2){\hat{\tau}_3\hat{\rho}_1\hat{\sigma}_0}-M(\gamma_0^2+\gamma_\pi^2){\hat{\tau}_1\hat{\rho}_0\hat{\sigma}_3}\right].
\end{equation}
We must reemphasize here that Eq. (\ref{CompactDia}) represents an approximate expression for the scattering matrix: although it takes a full account of the formation of the bound states. We have omitted terms which describe higher order interband scattering coherence effects that are proportional to $\gamma_0^n\gamma_\pi^m$ with $n,m>2$. 

\subsection{Multiple impurities}

At this point it will be instructive to briefly consider purely superconducting case first. The purpose is to highlight close analogy between the two-band model of $s^{\pm}$ SC with potential disorder, and an ordinary single band $s$-wave SC with magnetic impurities. Since for $M=0$ the structure of the $\hat{\cal T}$-matrix matches the one of the single-particle propagator, from Eq. (\ref{FullProp}) it follows that we can write 
\begin{equation}\label{Gpw}
\hat{G}(\bp,\omega_n)=\frac{-i\Omega_{\omega_n}{\hat{\tau}_0\hat{\rho}_0\hat{\sigma}_0}+\xi_\bp{\hat{\tau}_3\hat{\rho}_3\hat{\sigma}_0}+\Delta_{\omega_n}{\hat{\tau}_3\hat{\rho}_1\hat{\sigma}_0}}{\Omega_{\omega_n}^2+\xi_\bp^2+\Delta_{\omega_n}^2},
\end{equation}
where the renormalized Matsubara frequency $\Omega_{\omega_n}$ and the order parameter $\Delta_{\omega_n}$ must be determined self-consistently from
\begin{equation}\label{twntDLT}
\Omega_{\omega_n}=\omega_n+\Gamma_{\textrm{t}}\Omega_{\omega_n}\frac{\sqrt{\Omega_{\omega_n}^2+\Delta_{\omega_n}^2}}{\Omega_{\omega_n}^2+\epsilon_0^2 \Delta_{\omega_n}^2}, \quad
\Delta_{\omega_n}=\Delta+\Gamma_{\textrm{m}}\Delta_{\omega_n}\frac{\sqrt{\Omega_{\omega_n}^2+\Delta_{\omega_n}^2}}{\Omega_{\omega_n}^2+\epsilon_0^2 \Delta_{\omega_n}^2}.
\end{equation}
Here $\epsilon_0^2\equiv(1-J_s^2)/(1+J_s^2)$,  $\Gamma_{\textrm{t,m}}=\Gamma_0\pm\Gamma_\pi$ and $\Gamma_{0,\pi}=\pi \nu_Fn_{\textrm{imp}}|u_{0,\pi}|^2$ are the intraband and interband scattering rates. In what follows, we will assume, for simplicity, that the ratio $\Gamma_\pi/\Gamma_0$ is fixed. Curiously, impurity scattering makes an effective pairing field $\Delta_{\omega_n}$ to be dynamic, namely dependent on Matsubara frequencies, as it happens in the strong coupling approach of Eliashberg equations \cite{Eliashberg}. The analysis of the equation for the pair-potential $\Delta$ can be significantly simplified by introducing parameter ${\eta}_{\omega_n}=\Omega_{\omega_n}/\Delta_{\omega_n}$. In the limit of zero temperature, $T\to 0$, it attains the closed form 
\begin{equation}\label{Things2Solve}
\Delta\ln\left(\frac{\Delta}{\Delta_0}\right)=\int\limits_0^\infty d\omega_n
\left(\frac{1}{\sqrt{{\eta}_{\omega_n}^2+1}}-\frac{\Delta}{\sqrt{\omega_n^2+\Delta^2}}\right), \quad
{\eta}_{\omega_n}=\frac{\omega_n}{\Delta}+\left(\frac{1}{\tau_{\textrm{b}}\Delta}\right)\frac{{\eta}_{\omega_n}\sqrt{{\eta}_{\omega_n}^2+1}}{{\eta}_{\omega_n}^2+\epsilon_0^2},
\end{equation}
where $\Delta_0$ is the superconducting gap for a clean system, which can be expressed in terms of the coupling constant and cutoff energy via the relation $(\pi\nu_F\lambda_{\text{sc}})^{-1}=\ln(\omega_\Lambda/\Delta_0)$, and $\tau_{\textrm{b}}^{-1}=2\Gamma_\pi$. 
These equations match almost verbatim the corresponding equations obtained by Rusinov for the superconductor contaminated with paramagnetic impurities \cite{Rusinov}.  

\begin{figure}
\centering
\includegraphics[scale=0.35]{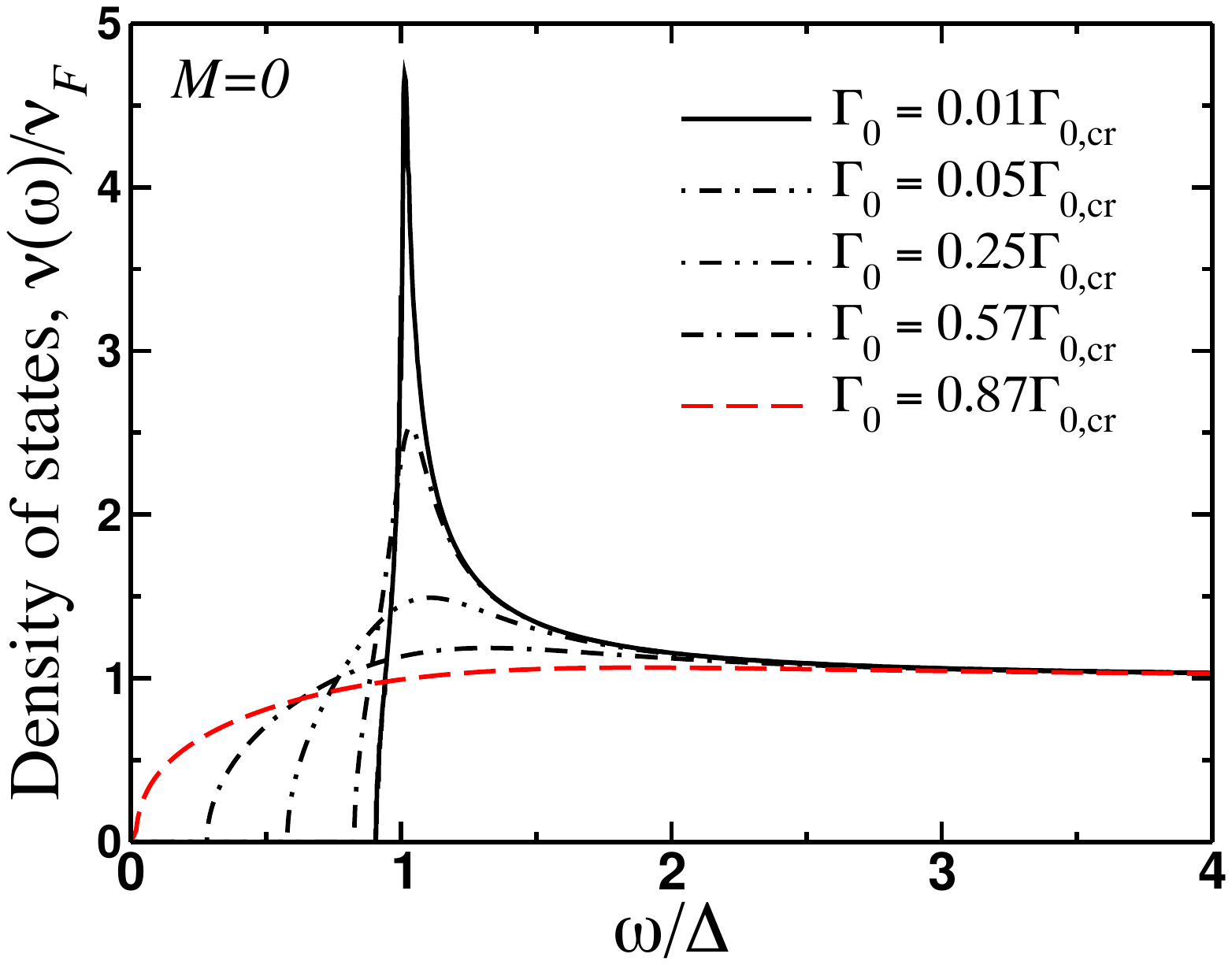}
\caption{Single-particle density of states as a function of energy computed in the superconducting state, Eq. (\ref{DOSM0}). We have used the following parameters: $\gamma_0=0.95$, $\gamma_\pi=0.175\gamma_0$, $\Gamma_\pi=0.175\Gamma_0$, $\epsilon_0=0.97$. When
  $\Gamma_0=\Gamma_{0,\textrm{cr}}\approx 1.43\Delta_0$ superconductivity becomes fully suppressed. The gapless superconductivity appears first for $\Gamma_0\approx 0.87\Gamma_{0,\textrm{cr}}$.}
\label{Fig-DoS-MO}
\end{figure}

\section{Density of states}\label{sec:dos}

\subsection{DoS in the superconducting state}

The density of states (DOS) in the case of $M=0$ can be computed using 
\begin{equation}\label{DOSM0}
\nu(\omega)\vert_{M=0}=\nu_F\textrm{Im}\left[\frac{\eta_\omega}{\sqrt{1-\eta_\omega^2}}\right],
\end{equation}
together with the first expression from Eq. (\ref{twntDLT}) and second expression from Eq. \eqref{Things2Solve}, and by performing analytical continuation from the set of discrete frequencies to the real axis of energies $i\omega_n\to\omega$ and $\eta_{\omega_n}\to-i\eta_{\omega}$. From Eq. (\ref{Things2Solve}) it is clear that the calculation of the DOS reduces to the problem of finding zeroes of the polynomial of sixth degree. Out of six roots, the physically relevant roots are a pair of complex conjugated ones. The results of the calculation of the DOS are shown in Fig. \ref{Fig-DoS-MO}. In agreement with the earlier studies \cite{VC2011,FVC2012,Dzero2021,Dzero2015}, we find that gapless superconductivity appears when the intraband scattering rate satisfies $0.87\Gamma_{0,\textrm{cr}}\leq\Gamma_0<\Gamma_{0,\textrm{cr}}$.

\begin{figure}
\centering
\includegraphics[scale=0.275]{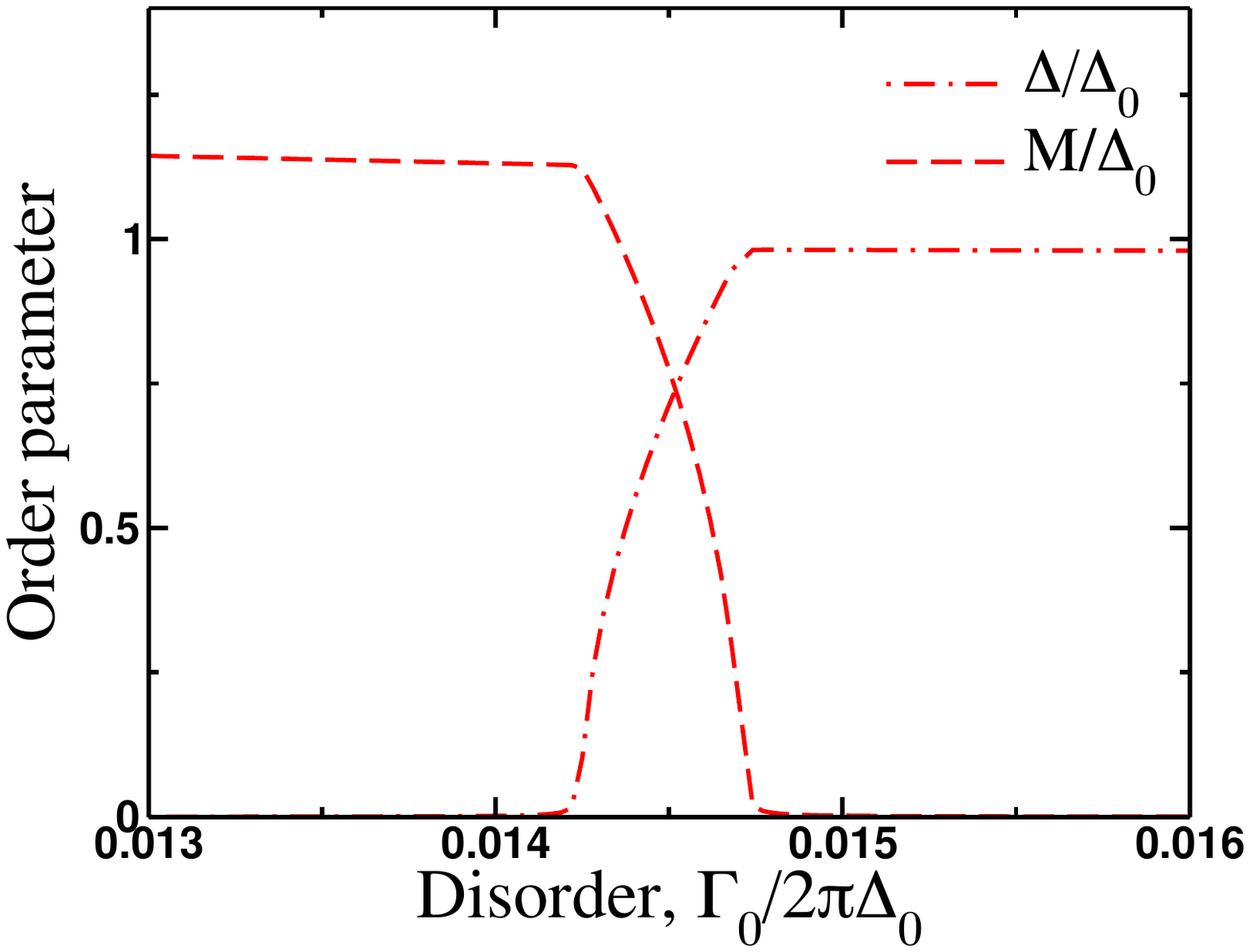}
\includegraphics[scale=0.275]{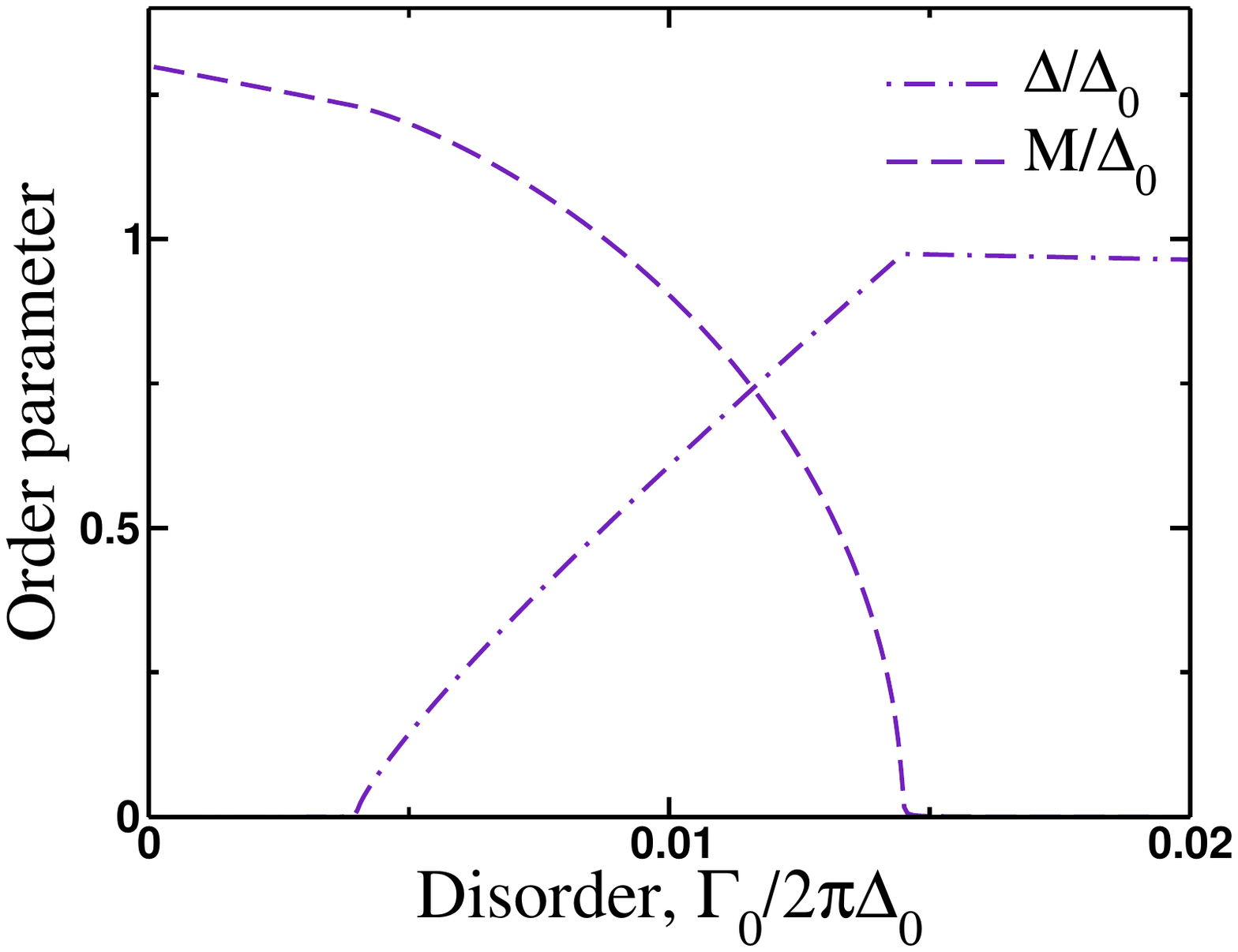}
\caption{Solution of the self-consistency equations \eqref{eq:Delta} and \eqref{eq:M} at $T=0$ with the single-particle propagator determined from Eq. (\ref{FullProp}). Superconductivity and spin-density-wave orders coexist in the narrow region of disorder concentration due to violation of the Anderson theorem. On the plots we took parameters that correspond to the critical temperature for the SDW order transition in a clean system to be higher than the corresponding critical temperature of the superconducting transition, so that we chose $M_0=1.3\Delta_0$. On the left panel $\gamma_0=0.12$, while on the right panel $\gamma_0=0.75$. The plots in both panels were done for $\gamma_\pi=0.175\gamma_0$ and $\Gamma_\pi=0.175\Gamma_0$.}
\label{Fig-OP}
\end{figure}

\subsection{DoS in the SC-SDW coexistence region}

We now turn our attention to the most interesting situation of the coexistence when both order parameters are simultaneously nonzero. For simplicity, we restrict our considerations to the limit of $T\to 0$ and solve the self-consistency equations for the order parameters as a function of scattering rate $\Gamma_0$. By reabsorbing the cutoff and coupling constant into the order parameter of the clean system, $(\pi\nu_F\lambda_{\text{m}})^{-1}=\ln(\omega_\Lambda/M_0)$, Eqs. \eqref{Eq2} and \eqref{Eq3} can be brought to the form 
\begin{align}
\Delta\ln\left(\frac{\Delta}{\Delta_0}\right)=\int\limits^{\infty}_{0}d\omega_n\left(\frac{\Delta_{\omega_n}}{\sqrt{\Omega^2_{\omega_n}+\Delta^2_{\omega_n}+M^2_{\omega_n}}}-\frac{\Delta}{\sqrt{\omega^2_n+\Delta^2}}\right), \label{eq:Delta} \\ 
M\ln\left(\frac{M}{M_0}\right)=\int\limits^{\infty}_{0}d\omega_n\left(\frac{M_{\omega_n}}{\sqrt{\Omega^2_{\omega_n}+\Delta^2_{\omega_n}+M^2_{\omega_n}}}-\frac{M}{\sqrt{\omega^2_n+M^2}}\right).\label{eq:M}
\end{align}
In these equations, and in analogy to the previous case of impure superconductivity, we introduced following modified notations 
\begin{align}\label{NewRelsSCSDW}
\Omega_{\omega_n}=\omega_n+\Gamma_{\textrm{t}}\Phi_{\omega_n}, \quad
\Delta_{\omega_n}=\Delta+\Gamma_{\textrm{m}}\Delta_{\omega_n}\frac{\Phi_{\omega_n}}{\Omega_{\omega_n}}, \quad
M_{\omega_n}=M-\Gamma_{\textrm{t}}M_{\omega_n}\frac{\Phi_{\omega_n}}{\Omega_{\omega_n}},
\end{align}
that are expressed though the additional function
\begin{equation}
\Phi_{\omega_n}=\Omega_{\omega_n}\frac{\sqrt{\Omega_{\omega_n}^2+\Delta_{\omega_n}^2+M_{\omega_n}^2}}{\Omega_{\omega_n}^2+C_{\omega_n}^2}, 
\end{equation}
where $C_{\omega_n}$ was obtained from the coefficient $c_1 $ in Eq. (\ref{BoundStates}) by replacing $\Delta\to\Delta_{\omega_n}$ and $M\to M_{\omega_n}$. Note that the minus sign in the expression for $M_{\omega_n}$ implies the absence of the Anderson theorem for the SDW order. The results of the numerical calculation for the phase diagram are shown in Fig. \ref{Fig-OP}.

\begin{figure}
\centering
\includegraphics[scale=0.19]{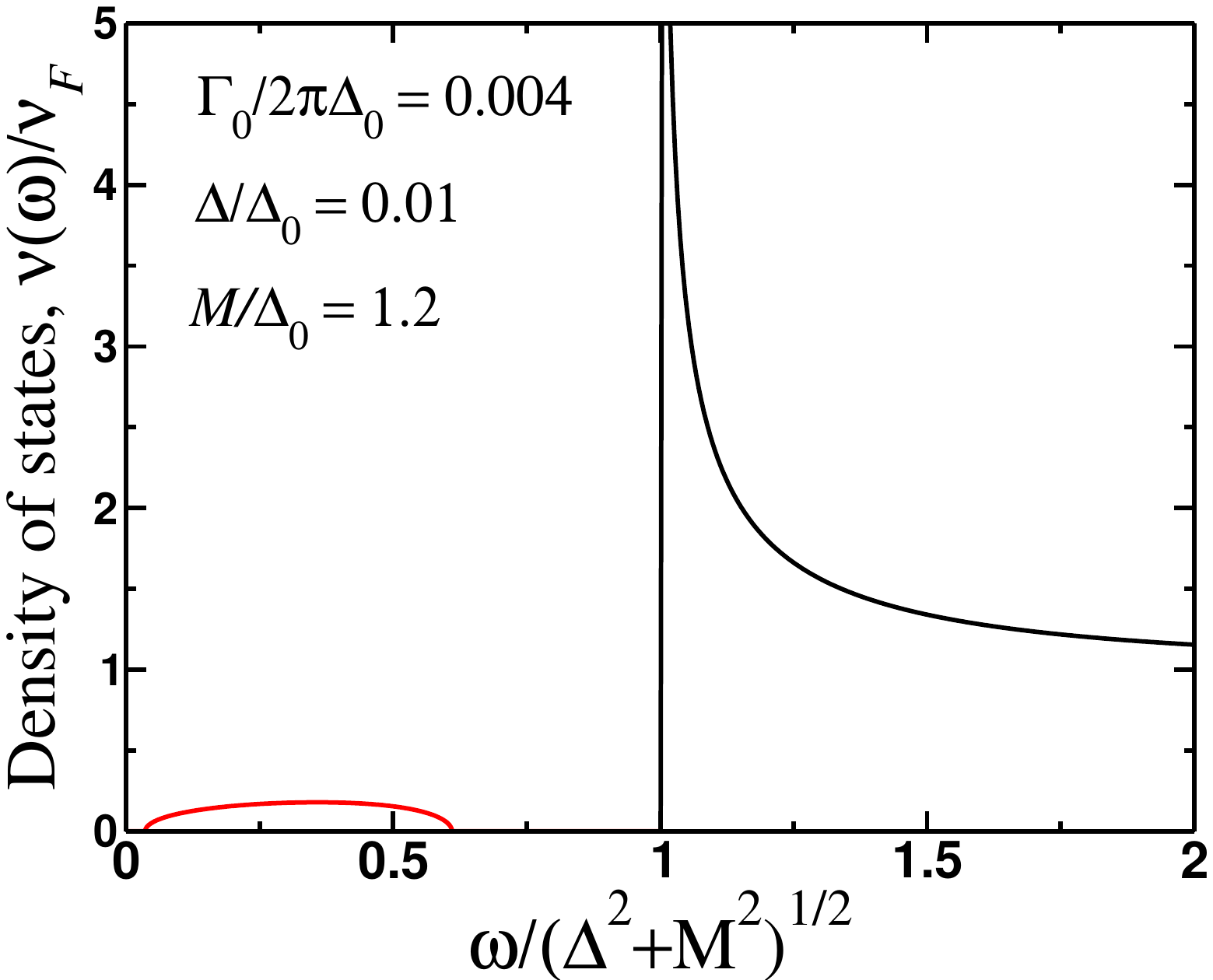}
\includegraphics[scale=0.19]{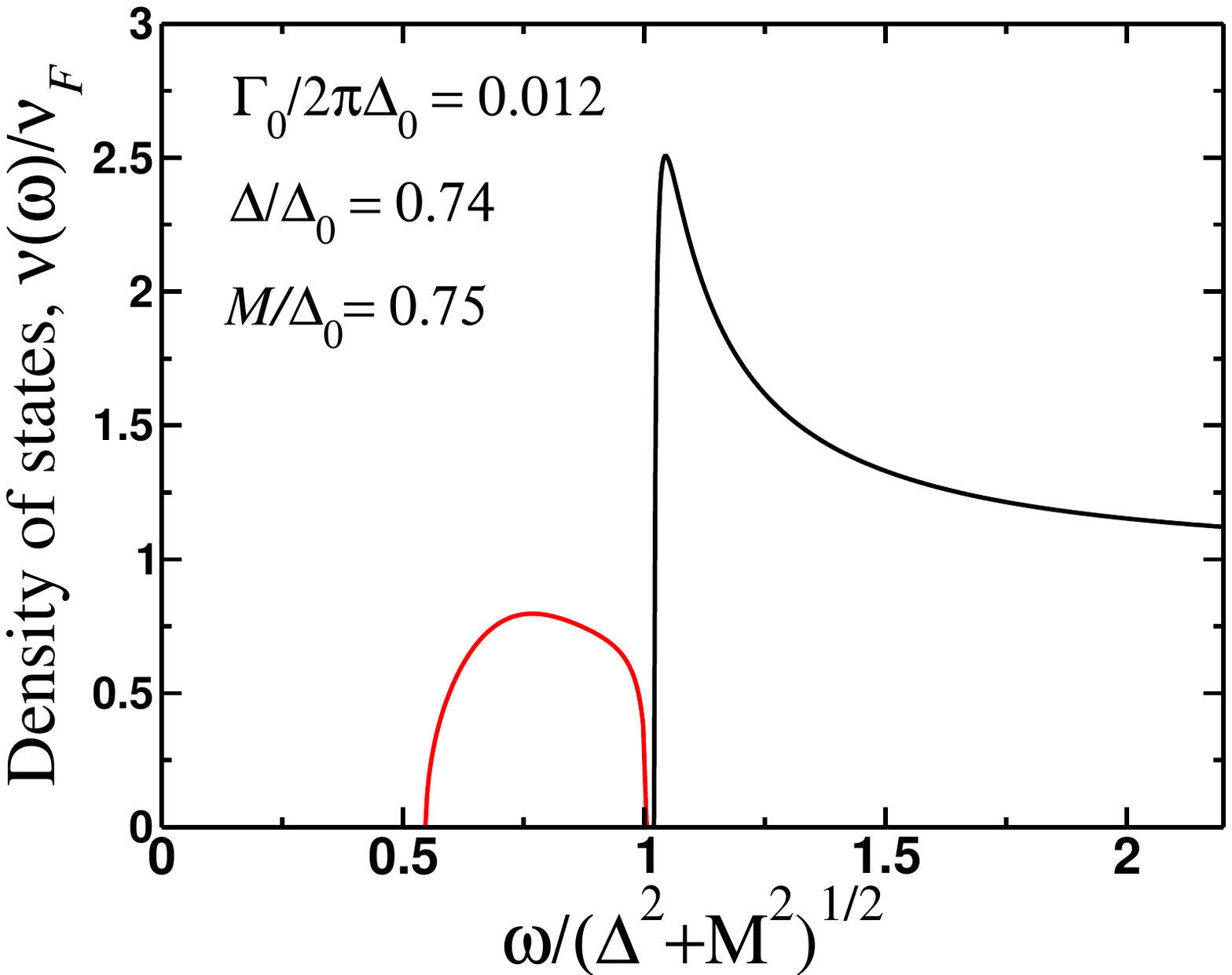}
\includegraphics[scale=0.19]{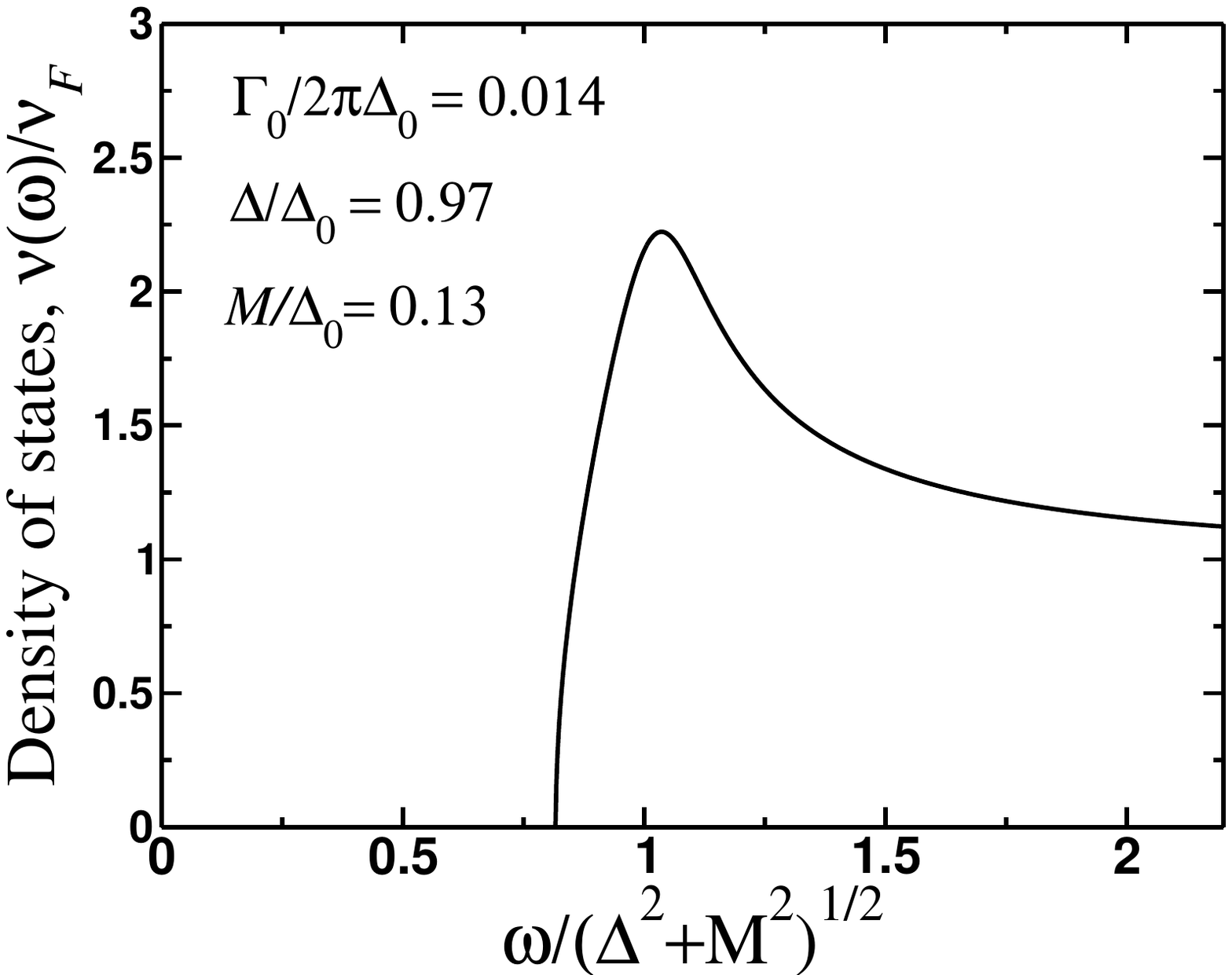}
\caption{Evolution of the single-particle density of states across the coexistence region (see Fig. \ref{Fig-OP}). In the region where $M\gg\Delta$, there is a wide impurity band with somewhat shallow structure. When $M\sim \Delta$ the width of the impurity band remains approximately the same while the value of the impurity DOS increases. Lastly, as $\Delta\gg M$, the impurity band merges with the DOS at $\omega\sim\Delta$. The model parameters are the same as in Fig. \ref{Fig-OP}.}
\label{Fig-DOS}
\end{figure}

After an analytical continuation, the dependence of the density of states on energy is given by
\begin{equation}\label{FullDOS}
\nu(\omega)=\nu_F\textrm{Im}\left[\frac{\Omega}{\sqrt{\Delta_\omega^2+M_\omega^2-\Omega^2}}\right].
\end{equation}
Here we have $\Omega=\omega+\Gamma_{\textrm{t}}{\Phi}_{\omega}$ and
\begin{equation}\label{tphiw1}
{\Phi}_{\omega}=\frac{\eta_\Delta\eta_M\sqrt{\eta_\Delta^2+\eta_M^2-\eta_\Delta^2\eta_M^2}}{\epsilon_0^2\eta_M^2+\epsilon_\textrm{m}^2\eta_\Delta^2-\eta_\Delta^2\eta_M^2}, \quad 
\eta_\Delta=\frac{\omega}{\Delta}+\frac{2\Gamma_{\pi}}{\Delta}{\Phi}_{\omega}, 
\quad \eta_M=\frac{\omega}{M}+\frac{2\Gamma_{\textrm{t}}}{M}\Phi_{\omega},
\end{equation}
with $\epsilon_\textrm{m}^2\approx(1-\gamma_0^2)/(1+\gamma_0^2)$. It is easy to see that, for example, in the limit $M\to 0$ ($\eta_M\to\infty$) we immediately recover the corresponding expression [second equation in (\ref{Things2Solve})] for the purely superconducting state.

We now need to solve these equations to find ${\Phi}_{\omega}$. Since we are interested in elucidating the contribution to the density of states from the impurity band, we have to find the complex roots of Eq. (\ref{tphiw1}). To do that, this equation must be recast into the polynomial form and solved numerically. Elementary power counting shows that finding all the roots of (\ref{tphiw1}) is equivalent to solving 
\begin{equation}\label{poly10}
\sum\limits_{n=0}^{10}a_n{\Phi}_{\omega}^n=0.
\end{equation}
Here the expansion coefficients are $a_{10}=(4\Gamma_\pi\Gamma_{\textrm{t}})^4$, $a_9=8\omega(4\Gamma_{\pi}\Gamma_{\textrm{t}})^3(\Gamma_\pi+\Gamma_{\textrm{t}})$, ..., $a_0=\omega^6(\omega^2-\Delta^2-M^2)$. 

Although the full analysis of the roots of (\ref{poly10}) can only be performed numerically, we can certainly obtain the analytical results for some combination of values of $\omega$, $M$ and $\Delta$. The simplest case to analyze is the region of the disorder concentrations for $\Delta\approx M$ and for small frequencies $\omega\ll\textrm{min}\{\Delta,M\}$. The latter condition is important for it would allow us to check whether there exists the single particle excitation threshold. In this limit the polynomial reduces to the one of degree four. In this case we find that all four roots are real and their corresponding contribution to the density of states is zero. This means that in the coexistence region superconductivity remains fully gapped. Our results for the density of states in the coexistence region found from the numerical solution of (\ref{poly10}) are shown in Fig. \ref{Fig-DOS}. As we have expected in the region of the phase diagram (Fig. \ref{Fig-DOS}) where $M\gg\Delta$, there is an impurity band at small $\omega$. As $\Gamma_0$ increases, the center of the impurity band moves to higher frequencies. 

\begin{figure}
\centering
\includegraphics[scale=0.55]{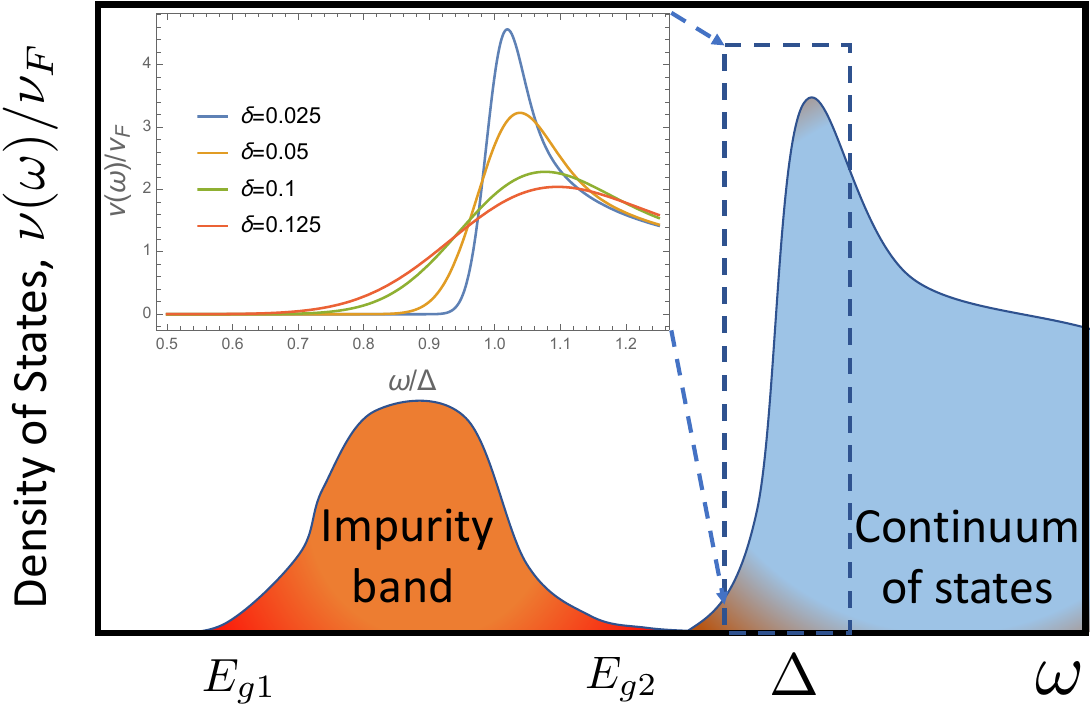}
\caption{Schematic plot for the density of states described beyond the mean field analysis of strong impurities. An account for the long-range spatial inhomogeneities leads to appearance of the Lifshitz-type tail states \cite{Lamacraft2001,Simons2001,Meyer2001,Ioffe2005,Feigelman2012,Fominov2016} extending from the sharp gaps of the spectral edge $\Delta$ and impurity band edges marked by $E_{g1,2}$. Inset shows the tail states extending from the BCS coherence peak calculated for the assumed spatial disorder with the Gaussian probability density of the order parameter as defined by Eq. \eqref{eq:DOS-LO-average} per Ref. \cite{LO}.}
\label{Fig-DoS-tails}
\end{figure}

\section{Discussion}\label{sec:discussion}

Thus far we have discussed how the singe-particle DOS changes if one goes beyond the SCBA. In this section we briefly touch on another interesting aspect of the problem relevant for disordered superconductors, namely the impact of spatial inhomogeneities with large correlation radius $r_c$ that exceeds both coherence length for the magnetic and superconducting orderings $r_c\gg \{\xi_\Delta,\xi_M\}$. The easiest way to model such system is to assume that coupling constants $\lambda_{\text{sc}}(\mathbf{r})$ and $\lambda_{\text{m}}(\mathbf{r})$ are now random in space and described by a certain correlation function $\langle \lambda(\mathbf{r})\lambda(\mathbf{r}')\rangle=F(|\mathbf{r}-\mathbf{r}'|/r_c)$. The exact form of this function is not important, it can be taken as a Gaussian, as long as spatial extend of this function gives the largest scale in the problem. Furthermore, this function may differ for $\lambda_{\text{sc}}(\mathbf{r})$ and $\lambda_{\text{m}}(\mathbf{r})$ correlations. In Ref. \cite{Dzero2021}, building on the original considerations of Larkin and Ovchinnikov \cite{LO} introduced for the conventional BCS superconductors, it was shown that in a model of two-band $s^{\pm}$-wave superconductor the spatial inhomogeneities lead to the broadening of the coexistence region between SDW order and superconductivity. In the conventional superconductor contaminated with long-range disorder, which produces potential (i.e. time-reversal-invariant) scattering only, inhomogeneities may lead to the smearing of the square-root anomaly near the threshold frequency of the coherence peak \cite{LO}. The smearing results in the shift of the peak and tail states going into the sub-gap region. We expect similar features to appear in the smearing of DOS near hard gaps of the impurity band. This picture is schematicaly illustrated in Fig. \ref{Fig-DoS-tails}.    

The qualitative physical picture that explains these features is most simply understood in the single component system (e.g. SC without SDW) but the same reasoning applies to the general situation. If the disorder correlation radius exceeds the length scale of superconductivity, this means that the  systems adjusts to the local (random) value of the order parameter $\Delta(\mathbf{r})$. Therefore, locally it is given by BCS expression in the clean limit. The global spectrum then may be found by averaging local DOS over the disorder realization of random $\Delta$
\begin{equation}\label{eq:DOS-LO}
    \nu(\omega)=\int\nu(\omega,\Delta)P(\Delta)d\Delta
\end{equation}
For instance, for the Gaussian probability density 
\begin{equation}
P(\Delta)=\frac{1}{\sqrt{2\pi\langle\delta\Delta^2\rangle}}\exp[-\delta\Delta^2/4\langle\delta\Delta^2\rangle], \quad \delta\Delta=\Delta-\langle\Delta\rangle,
\end{equation}
which is characterized by the strength of gap fluctuations with the average square of  $\langle\delta\Delta^2\rangle$, the average in Eq. \eqref{eq:DOS-LO} gives a universal curve near the spectral edge 
\begin{equation}\label{eq:DOS-LO-average}
\frac{\nu(\omega)}{\nu_F}=\frac{1}{2\sqrt{\delta}}D_{-1/2}\left(\frac{1-\varpi}{\delta}\right)\exp\left[-\frac{(1-\varpi)^2}{4\delta^2}\right],\quad \varpi=\omega/\langle\Delta\rangle,
\end{equation}
parametrized by a single dimensional quantity $\delta=\sqrt{\langle\delta\Delta^2\rangle}/\langle\Delta\rangle$. Here $D_n(z)$ is the parabolic-cylinder function, and the exact shape of DOS near the spectral edge is plotted as inset to Fig. \ref{Fig-DoS-tails} for different values of $\delta$.  In the extension of this picture to the coexistence scenario, the key thing to notice, is that if not for point-like disorder, it is only the combination of $\Delta^2+M^2$ that enters DOS expression. Therefore, one could invoke the same argument for the joint probability of spectral gap $\sqrt{\Delta^2+M^2}$. Treating short-range and long-range disorder on equal footing is a challenging task, however we expect the general picture with smeared gap edges to apply. Indirectly, this can be justified by an independent instanton calculus of tail states in SC with magnetic disorder \cite{Lamacraft2001,Simons2001,Meyer2001,Ioffe2005,Feigelman2012,Fominov2016} leading to the DOS structure consistent with that depicted in Fig. \ref{Fig-DoS-tails}. We only note that the precise energy dependence of the tails from each side of the impurity band needs to be reexamined in our model as it was shown earlier that the details are sensitive to the specifics of the model and the mechanism responsible for the fluctuations (e.g. fluctuations of the concentration of magnetic impurities and/or mesoscopic fluctuations of potential disorder). We close this section by pointing out that impurity bands may have significant effect on the thermodynamic properties, such as temperature dependence of the London penetration depth \cite{Prozorov2009,Prozorov2010}, as well as kinetic coefficients, such frequency dependent impedance \cite{Fominov2010,Kharitonov2012}. Experimentally these bands can be probed by the scanning tunneling techniques.   

\section{Acknowledgments}
We would like to thank Eugene Demler for useful discussion that stimulated this project. This work was financially supported by the National Science Foundation Grants No. DMR-2002795 (M. D.) and No. DMR-2203411 (A. L.). In part this work was performed at the Aspen Center for Physics, which is financially supported by the National Science Foundation Grant No. PHY-1607611. 

\bibliography{biblio}

\end{document}